# Extended Abstract: TRB Paper # 18-05804

# Intersection Approach Advisory Through Vehicle-to-Infrastructure Communication Using Signal Phase and Timing Information at Signalized Intersection

Md Salman Ahmed
Researcher, Center for Transportation Research, The University of Tennessee &
Vehicular Network Laboratory, East Tennessee State University
ahmedm@etsu.edu

Mohammad A Hoque, PhD
Director, Vehicular Network Laboratory, East Tennessee State University &
Researcher, Center for Transportation Research, The University of Tennessee
hoquem@etsu.edu

Asad J. Khattak, Ph.D.
Beaman Professor, Department of Civil & Environmental Engineering
The University of Tennessee, Knoxville, TN
akhattak@utk.edu

Total Words: 1084

**ACKNOWLEDGEMENTS**
This paper is based upon work supported by the US National Science Foundation under grant No. 1538139. The US Department of Transportation, Knoxville Traffic Division, Johnson City Traffic Division, and Precision Traffic provided additional support. Any opinions, findings, and conclusions or recommendations expressed in this paper are those of the authors and do not necessarily reflect the views of the sponsors.

August, 2017

Submitted for Presentation Only to the 97th Annual Meeting
Transportation Research Board
January 2018
Washington, D.C.

**INTRODUCTION**

Erratic acceleration and deceleration of vehicles while approaching traffic signals can potentially compromise traffic safety leading to crashes. Collisions that occur at intersections are more likely to be severe, on average, than non-intersection collisions. Moreover, the extent of fuel consumption and $CO_2$ emission becomes high due to frequent acceleration and deceleration at intersections. These problems can be ameliorated using assistive applications that can provide prior knowledge of traffic status and signals to drivers. Intelligent transportation systems (ITS), aided by connected vehicle technology, offer a plethora safety critical and assistive applications *(1–4)*. Over the years, researchers have approached different systems: such as optimal speed suggestions, safe speed for smooth driving, and signal predictions for the next phase. Some of these systems use cellular network (3G/4G/LTE) as a vehicle to infrastructure (V2I) communication and very few use the dedicated short-range communication (DSRC). However, in existing literature *(5–9)*, important system parameters and underlined challenges are not considered extensively. This paper explores an advisory system that can provide drivers traffic signal information along with speed recommendations to reach the destination intersection and analyzes the computational challenges associated with such a system. The system uses real-time signal phase and timing (SPaT) information for advisory calculation and DSRC for vehicle-to-infrastructure (V2I) communication. The design goal of the advisory application is to help drivers make informed decisions following audio and visual recommendations presented in a less distractive way.

**METHODOLOGY**

The proposed system uses a traffic controller, road side unit (RSU), on board unit (OBU) for the advisory calculations and dissemination through smart phones to provide optimal speed recommendations. This study analyzes challenges related to implementing an advisory algorithm considering important communication and computing parameters. Using raw SPaT data in different controllers is a challenging issue. Individual communication protocols are needed to decode and encode raw SPaT information as each traffic controller has its own data structure. Raw SPaT data is also vulnerable to security threats; if not encrypted, SPaT can be misused by intruders before reaching destination. However, encryption of SPaT is not within the scope of this research. To solve the issue of diverse SPaT data structure based on different controllers, this study compared SPaT data from two controllers (Siemens M60 ATC and Trafficware series 900 ATC) in hexadecimal format and then converted them to human readable format. The comparison led us to create a common decoding platform using a list of APIs. Because of the API implementation, it is possible to add more traffic controllers, as needed. A Road side unit (RSU) generates a string from decoded human readable SPaT format. The string has predefined fields and these fields can be expanded if required.

Mapping signal phase data based on legs of intersections becomes complicated if the number of phases exceeds 8 and includes separate pedestrian phases. Moreover, traffic signal phase design of an intersection might vary. The current implementation supports up to 8 phases and uses a triangular zone-setup configuration file to map phases to corresponding lanes. The configuration file consists of GPS coordinates defining the road geometry for each approach corresponding to individual phases.

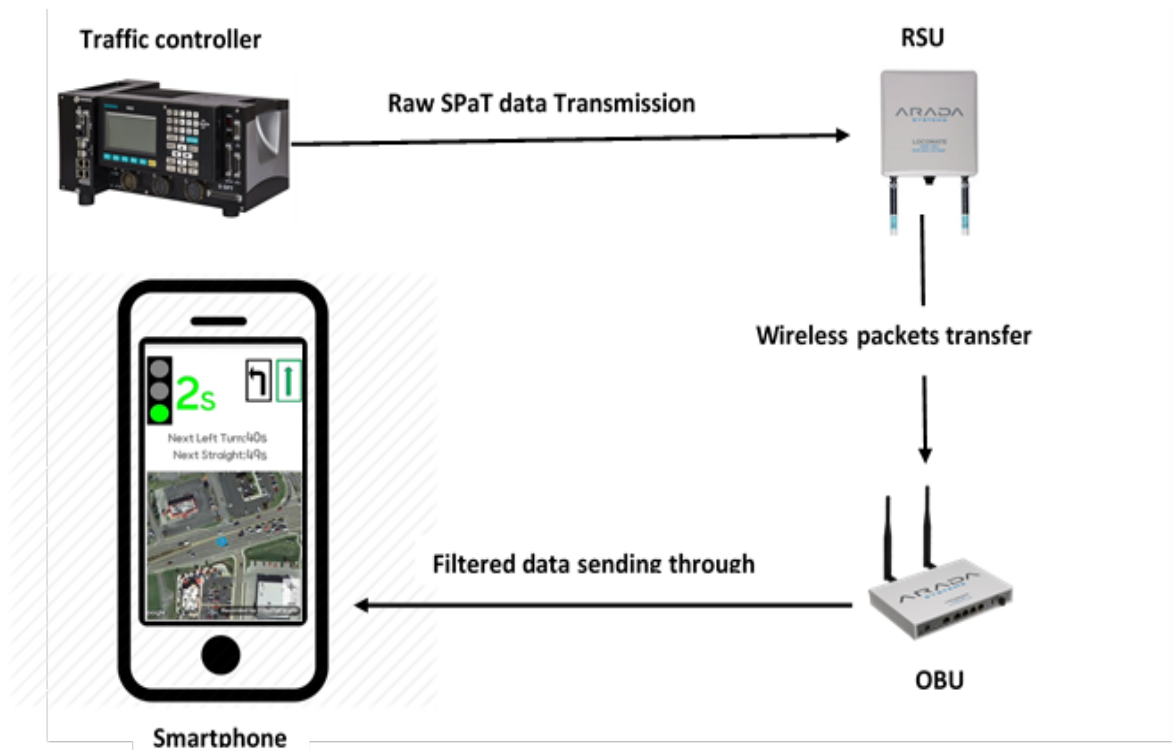

Figure 1: Overall System Architecture for signal advisory system

Figure 1 illustrates how the overall system works. The advisory system uses traffic controller to transmit raw SPaT data to RSU while RSU continuously scans for SPaT and reads data every tenth of a second. RSU packages decoded strings of raw SPaT along with zone information and transmit packages as wireless packets. The OBU unpacks these packages upon receipt. The OBU retrieves zone information from these packets, calculates a triangular area from the zonal information and compares its own coordinates with this zone. OBU filters out all other zones information from a packet if its coordinate falls into a particular zone. Filtered data is then sent to smartphone device using Bluetooth connectivity. Smartphone stores the data collected from OBU to compute the advisory algorithm. The advisory algorithm considers parameters like advisory start time, advisory end time, speed profile, fuel efficiency, visual message, and auditory notification. From the stored data combined with a vehicle's position, the algorithm calculates the vehicle's distance from an intersection, time for phase change and plots the position in a Google map. The advisory algorithm notifies a user about the time left for a green phase to turn red or a red phase to turn green. To reduce confusion during a driver's decision making, the algorithm considers maximum distance to start recommendation and minimum distance to stop recommendation. One feature of the advisory system is that the Smartphone display is designed in a way not to distract the driver. The display provides information about current phase; the time left for the current phase is displayed using a countdown timer. It also includes the time left for next phases. If the traffic signal changes, then the Smartphone can notify user both visually and audibly.

**FINDINGS**

With the help of city traffic division, the success of interface between a real-life traffic controller and RSU as well as communication between RSU and OBU was tested in Knoxville, TN at the intersection of Volunteer Blvd. and Andy Holt Tower Road. The SPaT advisory application was first tested in a lab environment based on the standard 8 phase intersection on North State of Franklin & West Market Street in Johnson City, TN. Although the proposed algorithm computes speed advisory, the graphical interface of this feature will be implemented in the future. The smartphone application calculates and displays countdown information about the phases, using DSRC. The field test proved that the application does not provide advisory unless a vehicle is located within the advisory display zone, which starts approximately 500 meters ahead of the intersection. Verification of phase timer was also conducted to examine if a green timer is followed by a yellow timer. A test case was made to examine if the system notifies a phase change and while the phase changes to green, the smartphone shows it visually and also audially by beeping. Overall, the intersection approach advisory system was successfully tested through I2V communication, based on SPaT information.

**CONCLUSIONS**

The primary focus of the study is to develop and analyze issues in implementing an intersection approach advisory system. DSRC equipment along with in-vehicle units were used for field testing. Advisories based on SPaT information were generated and transmitted to test vehicles and displayed through Smartphones to drivers. Simulations helped augment the field results. The experiments reported in this study can help improve their safety. We plan to further develop applications that also account for fuel consumption and delay reductions in the future.